\documentclass{ws-p9-75x6-50}

\begin{document}

\title{On Pauli's invention of non-Abelian Kaluza-Klein Theory in 1953}

\author{N. Straumann}

\address{
Institute of Theoretical Physics\\
           University of Z\"urich, Switzerland\\
E-mail: norbert@physik.unizh.ch}


\maketitle

\abstracts{There are documents which show that Wolfgang Pauli developed in 1953 the first consistent generalization of the five-dimensional theory of Kaluza, Klein, Fock and others to a higher dimensional internal space. Because he saw no way to give masses to the gauge bosons, he refrained from publishing his results formally.}

\section{Introduction}

As Chairman of this parallel session I use the opportunity to make a few opening remarks on a largely unknown chapter of the early history of Kaluza-Klein theories.\\
Already in summer 1953 W. Pauli developed in considerable detail the non-Abelian generalization of Kaluza-Klein theories. He described his attempt extensively in some letters to A. Pais, which are now published in Vol. IV, Part II of Pauli's collected letters\cite{kn:Pau1}, as well in two seminars in Z\"urich on November 16 and 23, 1953. The latter have been written up by Pauli's pupil P. Gulmanelli\cite{kn:Gul}. By specialization (independence of spinor fields on internal space) Pauli got all important formulae of Yang and Mills, as he later (Feb. 1954) pointed out in a letter to Yang\cite{kn:Pau6}, after a talk of Yang in Princeton. Pauli did not publish his study, because he was convinced that "one will always obtain vector mesons with rest mass zero" (Pauli to Pais, 6 Dec., 1953).

\section{The Pauli letters to Pais}

At the Lorentz-Kammerlingh Onnes conference in Leiden (22-27 June 1953) A. Pais talked about an attempt of describing nuclear forces based on isospin symmetry and baryon number conservation. In this contribution he introduced fields, which do not only depend on the spacetime coordinates $x$, but also on the coordinates $\omega$ of an internal isospin space. The isospin group acted, however, globally, i.e., in a spacetime-independent manner.\\
During the discussion following the talk by Pais, Pauli said:\\
\, \\
\textit{...I would like to ask in this connection whether the transformation group with constant phases can be amplified in a way analogous to the gauge group for electromagnetic potentials in such a way that the meson-nucleon interaction is connected with the amplified group...}\\
\, \\
Stimulated by this discussion, Pauli worked on the problem, and wrote on July 25, 1953 a long technical letter to Pais\cite{kn:Pau2}, with the motto: "Ad \textit{usum Delfini} only". This letter begins with a personal part in which Pauli says that "the whole note for you is of course written in order to drive you further into the real virgin-country". The note has the interesting title:\\
\, \\
\textit{Written down July 22-25 1953, in order to see how it looks. Meson-Nucleon Interaction and Differential Geometry.}\\
\, \\
In this manuscript, Pauli generalizes the original Kaluza-Klein theory to a six-dimensional space and arrives through dimensional reduction at the essentials of an $SU(2)$ gauge theory. The extra-dimensions form a two-sphere $S^2$ with space-time dependent metrics on which the $SU(2)$ operates in a space-time-dependent manner. Pauli enphasizes that this transformation group "seems to me therefore the \textsf{natural generalization of the gauge-group} in case of a two-dimensional spherical surface". He then develops in 'local language' the geometry of what we now call a fibre bundle with a homogeneous space as typical fiber (in this case $SU(2)/U(1)$).\\
Since it is somewhat difficult to understand exactly what Pauli did, we give some details, using more familiar formulations and notations\cite{kn:Strau}.\\
Pauli considers the six-dimensional total space $M \times S^2$, where $S^2$ is the two-sphere on which $SO(3)$ acts in the canonical manner. He distinguishes among the diffeomorphisms (coordinate transformations) those which leave $M$ pointwise fixed and induce space-time-dependent rotations on $S^2$:
\begin{equation}
(x,y)\rightarrow [x,R(x).y].
\end{equation}
Then Pauli postulates a metric on $M \times S^2$ that is supposed to satisfy three assumptions. These led him to what is now called the non-Abelian Kaluza-Klein ansatz: The metric $\hat{g}$ on the total space is constructed from a space-time metric $g$, the standard metric $\gamma$ on $S^2$, and a Lie-algebra-valued 1-form,
\begin{equation}
A=A^a T_a \, , \,  A^a=A^a_\mu dx^\mu,
\end{equation}
on $M$ ($T_a$, $a=1,2,3$, are the standard generators of the Lie algebra of $SO(3)$) as follows: If $K^i_a \partial / \partial y^i$ are the three Killing fields on $S^2$, then
\begin{equation}
\hat{g}=g-\gamma_{ij}[dy^i+K^i_a(y)A^a]\otimes[dy^j+K^j_a(y)A^a].
\end{equation}
In particular, the nondiagonal metric components are
\begin{equation}
\hat{g}_{\mu i}=A^a_\mu(x)\gamma_{ij}K^j_a.
\end{equation}
Pauli does not say that the coefficients of $A^a_\mu$ in Eq.(4) are the components of the three independent Killing fields. This is, however, his result, which he formulates in terms of homogeneous coordinates for $S^2$. He determines the transformation behavior of $A^a_\mu$ under the group (1) and finds in matrix notation what he calls "the generalization of the gauge group":
\begin{equation}
A_\mu \rightarrow R^{-1} A_\mu R+R^{-1}\partial_\mu R.
\end{equation}
With the help of $A_\mu$, he defines a covariant derivative, which is used to derive "field strengths" by applying a generalized curl to $A_\mu$. This is exactly the field strength that was later introduced by Yang and Mills. To our knowledge, apart from Klein's 1938 paper, it appears here for the first time. Pauli says that "this is the \textit{true} physical field, the analog of the \textit{field strength}" and he formulates what he considers to be his "main result":\\
\, \\
\textit{The vanishing of the field strength is necessary and sufficient for the $A^a_\mu (x)$ in the whole space to be transformable to zero.}\\
\, \\
It is somewhat astonishing that Pauli did not work out the Ricci scalar for $\hat{g}$ as for the Kaluza-Klein theory. One reason may be connected with his remark on the Kaluza-Klein theory in Note 23 of his relativity article\cite{kn:Pau3} concerning the five dimensional curvature scalar (p. 230):\\
\, \\
\textit{There is, however, no justification for the particular choice of the five-dimensional curvature scalar $P$ as integrand of the action integral, from the standpoint of the restricted group of the cylindrical metric (gauge group). The open problem of finding such a justification seems to point to an amplification of the transformation group.}\\
\, \\
In a second letter\cite{kn:Pau4}, Pauli also studies the dimensionally reduced Dirac equation and arrives at a mass operator that is closely related to the Dirac operator in internal space $(S^2,\gamma)$. The eigenvalues of the latter operator had been determined by him long before\cite{kn:Pau5}. Pauli concludes with the statement: "So this leads to some rather unphysical \textit{shadow particles}".\\
Pauli learned about the related work of Yang and Mills in late February, 1954, during a stay in Princeton, when Yang was invited by Oppenheimer to return to Princeton and give a seminar on his joint work with Mills. About this seminar Yang reports\cite{kn:Ya}: "Soon after my seminar began, when I had written down on the blackboard $(\partial_\mu -i \epsilon B_\mu )\Psi$, Pauli asked: \textit{What is the mass of this field $B_\mu$?}, I said we did not know...".\\
This problem was Pauli's main concern. He emphasized it repeatedly, most explicity in the second letter\cite{kn:Pau4} to Pais on December 6, 1953, after he had made some new calculations and had given the two seminar lectures already mentioned. He adds to the Lagrangian what we now call the Yang-Mills term for the field strengths and says that "one will always obtain vector mesons with rest-mass zero (and the rest-mass if at all finite, will always remain zero by all interactions with nucleons permitting the gauge group)." To this Pauli adds: "One could try to find other mesons fields", and he mentions, in particular, the scalar fields which appear in the dimensional reduction of the higher-dimensional metric. In view of the Higgs mechanism this is an interesting remark.\\
In a letter to Yang\cite{kn:Pau6} shortly after Yang's Princeton seminar, Pauli repeats: "But I was and still am disgusted and discouraged of the vector field corresponding to particles with zero rest-mass (I do not take your excuses for it with 'complications' seriously) and the difficulty with the group due to the distinction of the electromagnetic field remains." Formally, Pauli had, however, all important equations, as he shows in detail, and he concludes the letter with the sentence: "On the other hand you see, that your equations can easily be generalized to include the $\omega$-space" (the internal space). As already mentioned, the technical details have been written up by Pauli's pupil P. Gulmanelli\cite{kn:Gul}. It would be nice if someone would translate these seminar notes from Italian to English.\\
Shortly after the $100^{th}$ anniversary of Wolfgang Pauli I thought that it would be fitting to tell you (as a former student of Pauli) this interesting story.

\end{document}